\documentclass[preprintnumbers,showpacs,amsmath,amssymb,floatfix,9pt,prd,twocolumn,
superscriptaddress,nofootinbib]{revtex4}
\usepackage{latexsym}
\usepackage{epsfig}
\usepackage{amssymb}

\begin{document}

\title {The Dark Energy Star and Stability analysis}
\author{Piyali Bhar}
\email{piyalibhar90@gmail.com  } \affiliation{ {  Department of
Mathematics, Jadavpur University, Kolkata 700 032, West Bengal,
India}}

\author{{ Farook Rahaman}}
\email{rahaman@iucaa.ernet.in} \affiliation{ {  Department of
Mathematics, Jadavpur University, Kolkata 700 032, West Bengal,
India}}

\begin{abstract}

\textbf{Abstract:}   We have proposed a new model of dark energy
star consisting of five  zones namely, solid core of constant energy density, the thin shell
between core and interior,
    an inhomogeneous interior region with
anisotropic pressures, thin shell and the exterior vacuum region.
 We have discussed various physical properties. The model satisfies all
 the physical requirements. The stability condition under small linear
 perturbation has also been discussed.

\end{abstract}

\maketitle

\section{Introduction}
{\small The study of dark energy star has become a subject of
interest due to the fact that expansion of the universe is
accelerating which was suggested by High-z supernova Search Team
in $1998$ by observing type $1a$ supernova.
Dark energy is the most acceptable hypothesis to explain this accelerating expansion of the universe. According to the work done of Plank mission team and based on the standard model of cosmology the total mass energy of the universe contains $4.9$ percent ordinary matter, $26.8$ percent dark matter and $68.3$ percent dark energy. Dark matter is attractive in nature which can not seen by a telescope and it does not absorbs or emits light or any gravitational waves. But its existence has been proved by gravitational effects on visible matter and gravitational lensing of background radiation. On the other hand the dark energy needs to have a strong negative pressure in order to explain the rate of  accelerating expansion of the universe.\\

 ~~~~To construct a model of a relativistic star we generally assume that the underlying fluid distribution is homogeneous and isotropic. But it is proved by advance researches that the highly compact astrophysical objects like X-ray pulsar, Her-x-1, X-ray buster 4U 1820-30, millisecond pulsar SAXJ1804.4-3658 etc. whose density of core is expected to be beyond the nuclear density $(\sim 10^{15}gm/cc)$ shows the anisotropy. Anisotropy may occurs in the existence of solid core, in presence of type P superfluid, phase transition, rotation, magnetic field, mixture of two fluid, existence of external field etc. In case of anisotropy distribution the pressure inside the fluid sphere is not homogeneous in nature, it can be decomposed into two parts radial pressure $p_r$ and the transverse pressure $p_t$. So obviously $p_r\neq p_t$. Where $p_t$ is in the orthogonal direction to $p_r$. $\Delta=p_t-p_r$ is defined as the anisotropic factor whereas $\frac{\Delta}{r}$ is defined as anisotropic force which is repulsive in nature if $p_t>p_r$ and attractive if $p_t<p_r$.\\

 ~~~In this paper,we are going to model of a anisotropic dark energy star characterized by the parameter  $\omega=\frac{p_r}{\rho}$, where $p_r$ and $\rho$ are respectively the radial pressure and energy density. For accelerating expansion the dark energy parameter $\omega<-\frac{1}{3}$ is required. $-1<\omega<-\frac{1}{3}$ is referred to as quintessence. The region where $\omega<-1$ is named as phantom regime which has a peculiar property namely infinitely increasing energy density. $ \omega=-1$ corresponds to Einstein cosmological constant and this value is called cosmological constant barrier or phantom divide.

 ~~~~~A two dimensional Brans-Dicke star model with exotic matter and dark energy was studied in \cite{yanjun}. In that paper,  the author has taken the matter state equation as $p=\gamma \rho$, where $\gamma$ is the state parameter of exotic matter which satisfies $-\frac{1}{4}<\gamma<0$ and has shown that the mass of the star decrease if $\gamma$ decrease. Anisotropic dark energy star has been discussed in \cite{Chan1}. Star model with dark energy has been proposed in \cite{Chan2}. In this paper the authors have proposed a model of dark energy star consisting of four region and by analyzing the model they conclude that for static solution at least one of the regions must be constituted by dark energy. Anisotropic dark energy star was studied by Ghezzi {\it et al} \cite{Ghezzi}. In this paper the authors have assumed variable dark energy  which suffers a phase transition at a critical density and the anisotropy. The anisotropy is concentrated on a thin shell where the phase transition takes place, while the rest of the star remains isotropic. The solutions shows several features similar to the gravastar model. Lobo\cite{lobo1} has given a model of stable dark energy star by assuming two spatial type of mass function one is of constant energy density and the other mass function is Tolman-Whitker mass. All the features of the dark energy star has been discussed and the system is stable under small linear perturbation. The van der Waals quintessence stars have been studied in \cite{lobo2}. In that work, the construction of inhomogeneous compact spheres supported by a van der Waals equation of state is explored. van der Waals gravastar, van der Waals  wormhole have also been discussed. Variable Equation of State for Generalised Dark Energy Model has been studied in \cite{saibal}. Yadav \emph{et al.}have given a dark energy models with variable equation of state parameter in \cite{jadav}. Some other works on dark energy star are in \cite{neil,dubravko,wen,paul,vlad,jorge,chris,bridget}.\\

 ~~~The plan of the paper is  as follows: In section II basic field equations have been given. The model of dark energy star,exterior spacetime and junction condition, TOV equation,Energy condition, Mass-radius relation have been respectively discussed in section III-VII. The stability analysis under small radial perturbation has been studied in section VIII. Finally in section IX we have provided a short discussion and made some concluding remarks.

\section{Basic Field Equations}
A static and spherically symmetry spacetime in curvature coordinates is given by the following metric

\[ds^{2}=-exp\left[{-2\int_r^{\infty}g(\tilde{r})d\tilde{r}}\right]dt^{2}+\frac{dr^{2}}{1-\frac{2m}{r}}\]
\begin{equation}~~~~~~~~~~~~~~~~~~~~~~+r^{2}(d\theta^{2}+\sin^{2}\theta d\phi^{2}),
\end{equation}
where $g(r)$ and $m(r)$ are arbitrary functions of the radial parameters r. The function $m(r)$ is the quasi local mass and is denoted as the mass function.The factor $g(r)$ is termed as the 'gravity profile' which is used to measure the acceleration due to gravity by the relationship $\mathcal{A}=\sqrt{1-\frac{2m}{r}}g(r)$. For inward gravitational attraction $g(r)>0$ and $g(r)<0$ for outward gravitational repulsion. One can note that $\Phi(r)=-\int_r^{\infty}g(r)dr$, here $\Phi(r)$ is denoted as the redshift function.\\

The stress energy momentum tensor is given by the equation
\begin{equation}
T_{\mu\nu}=(\rho+p_t)U_{\mu}U_{\mu}+p_t g_{\mu\nu}+(p_r-p_t)\chi_{\mu}\chi_{\nu}
\end{equation}
where $U^{\mu}$ is the vector $4$-velocity,$\chi^{\mu}$ is the spacelike vector. $\rho(r)$ is the energy density and $p_r$ is the radial pressure measured in the direction of the spacelike vector. $p_t$ is the transverse pressure in the orthogonal direction to $p_r$ and $\Delta=p_t-p_r$ is called the anisotropic factor.\\

Using the Einstein field equations $G_{\mu\nu}=8\pi T_{\mu \nu}$ we get the following relationship,
\begin{equation}
m'=4\pi r^{2}\rho
\end{equation}
\begin{equation}
g=\frac{m+4\pi r^{3}p_r}{r(r-2m)}
\end{equation}
\begin{equation}
p_r'=-\frac{(\rho+p_r)(m+4\pi r^{3}p_r)}{r(r-2m)}+\frac{2}{r}(p_t-p_r)
\end{equation}
where $G_{\mu \nu}$ is the Einstein tensor and 'prime' denotes the derivative with respect to radial coordinate $r$.\\
The dark energy equation of state is given by the following equation
\begin{equation}
p_r=\omega \rho
\end{equation}
where $\omega <0$ is the equation of state parameter.\\

~~Now one can note that we have five unknown functions namely $\rho, p_r, p_t,m(r), g(r)$ and four equations[(3)-(6)]. To solve the set of equations let us assume a particular choice of the energy density $\rho$. This particular choice of $\rho$ was chosen earlier by Dev and Gleiser \cite{dev} to discuss anisotropic star model. Rahaman \emph{et al.}have also used this density function in\cite{fr1}. Using this particular choice of energy density we will find out the other parameters in explicit form.

\section{ Model of the dark energy Star }

Let us choose the energy density of the star as
\begin{equation}
\rho=\frac{1}{8\pi}\left(\frac{a}{r^{2}}+3b\right)
\end{equation}
where both $a$ and $b$ are constants, e.g $a=\frac{3}{7}$ and $b=0$ corresponds to relativistic
Fermi gas which can be seen in the ultradense  cores of a neutron star \cite{9}
 and for $a=\frac{3}{7}, b \neq 0$ we get relativistic fermi gas core in a
 constant density background.\\
Using $(7)$ into $(3)$ we obtain the expression of the mass function as,
\begin{equation}
m=\frac{1}{2}r(a+br^{2})
\end{equation}
Solving equation $(4)-(6)$ we get,
\begin{equation}
g(r)=\frac{a(1+\omega)+br^{2}(1+3\omega)}{2r(1-a-br^{2})},
\end{equation}
  From     Buchdahl limit,  $\frac{2m(r)}{r}<1$ one can see from
 equation (8) as $1-a-br^{2}>0$.   This provides   a constraint on
 the parameters $a,~b$ and radius r as
  \[r^{2}<\frac{1-a}{b}~~~~~~~~~~~~~~~~~~~~~~~~~~~~~~~~~~~~~~~~~~~~~~~~~~~~~~~~~~(9.1)\]
  This yields the restriction on $a,~b$ as
\[a<1 ~and~b>0.\]
   We have  chosen  $a=0.5$ and $b=0.00$1 motivated by the choice of
   Dev and Glaiser \cite{dev}.
   They chose $a=3/7$  and $b=0$ to describe their model.
    Our chosen values of $a$ and $b$ are very close to their choice.
    Now using those values for $a$ and $b$ from equation (9.1),
     we get $r^{2}<500$ i.e $r<22.36 $.
Now, $g(r)>0$
gives,\[\omega>-\frac{a+br^{2}}{a+3br^{2}}~~~~~~~~~~~~~~~~~~~~~~~~~~~~~~~~~~~~~~~~~~~~~~~~~~~~~~~~~~(9.2)\]
 For large r i.e. in the maximum limit,  we get by using (9.1) in (9.2) as
 \[\omega>-\frac{1}{3-2a}\]
 For small r, from (9.2) we have
 $\omega>-1$.\\
 Therefore, \[\omega>max\left(-1,-\frac{1}{3-2a}\right)~~~~~~~~~~~~~~~~~~~~~~~~~~~~~~~~~~~~~~~~~~~~~~(9.3)\]

 For the above choice, $a=0.5$,  $max\left(-1,-\frac{1}{3-2a}\right) = -0.5$ i.e. $\omega>-0.5$.

 Again for $g(r)<0$, we have
\[\omega<-\frac{a+br^{2}}{a+3br^{2}}~~~~~~~~~~~~~~~~~~~~~~~~~~~~~~~~~~~~~~~~~~~~~~~~~~~~~~~~~(9.4)\]
For small r, equation (9.4) yields\\
$\omega<-1$
and for large r i.e. in the maximum limit, \[\omega<  -\frac{1}{3-2a} \]
Thus,
\[\omega<min \left(-1,-\frac{1}{3-2a}\right)~~~~~~~~~~~~~~~~~~~~~~~~~~~~~~~~~~~~~~~~~~~(9.5)\]
For our choice of $a$,    we get  $ \omega<-1$.
So, we have to choose $\omega$ either $  \omega > -0.5 $  or  $ \omega<-1 $  for plot.\\

 The profiles of  g(r) have been plotted in $fig~1$ and
$fig~2$
 for $-0.45\leq \omega \leq -0.1$ and $\omega<-1$ respectively.\\
From the fig.~1 we see that $g(r)>0$ when $-0.45 \leq \omega \leq
-0.1$ and fig.~2 shows that for $\omega<-1$, $g(r)<0 $ for above
choice of $a$ and $b$.
\begin{figure}[htbp]
    \centering
        \includegraphics[scale=.35]{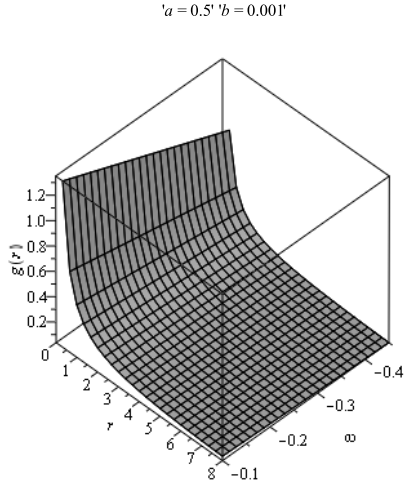}
       \caption{``gravity profile",~$g(r)$,~ has been plotted against $r$ when $-0.45\leq \omega \leq-0.1 $}
    \label{fig:3}
\end{figure}
The radial and transverse pressure can be obtained as,
\begin{equation}
p_r=\frac{\omega}{8\pi}\left(\frac{a}{r^{2}}+3b\right)
\end{equation}
\begin{equation}
p_t=\frac{(1+\omega)(a+3br^{2})}{32\pi r^{2}(1-a-br^{2})}\left[a(1+\omega)+br^{2}(1+3\omega)\right]+\frac{3b\omega}{8\pi}
\end{equation}
The matter density, radial and transverse pressures have been depicted in $fig~3.$\\
The anisotropy factor $\Delta$ is given by,
\begin{equation}
\Delta=\frac{(1+\omega)(a+3br^{2})}{32\pi r^{2}(1-a-br^{2})}
\left[a(1+\omega)+br^{2}(1+3\omega)\right]-\frac{\omega a}{8\pi r^{2}},
\end{equation}
which  has been shown in $fig.4$ and $fig.5$ respectively for
$-0.45 \leq \omega \leq -0.1$
 and $\omega<-1$. Now $\frac{\Delta}{r}$ re presents a force due to the pressure anisotropy.
 The force will be repulsive in nature i.e. in the outward direction if $p_t>p_r$ and attractive
  if $p_t<p_r$ or alternatively $\Delta<0$. For our stellar model configuration
  (see fig.4 and fig.5) $\Delta >0$ for both the cases when
   $-0.5<\omega<-0.1$ and   for phantom regime i.e. for   $\omega<-1$.
\begin{figure}[htbp]
    \centering
        \includegraphics[scale=.35]{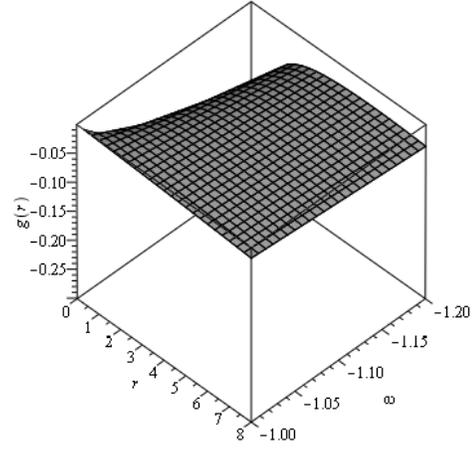}
       \caption{``gravity profile",~$g(r)$,~has been plotted against $r$ when $ \omega \leq -1 $, $a=0.45$ and $b=0.001$.}
    \label{fig:3}
\end{figure}
\begin{figure}[htbp]
    \centering
        \includegraphics[scale=.25]{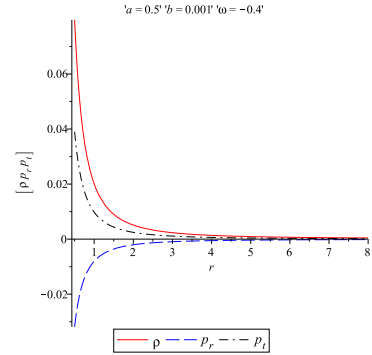}
       \caption{Matter density $\rho$,radial pressure $p_r$ and transverse pressure $p_t$ of the dark energy model has been plotted against $r$}
    \label{fig:3}
\end{figure}
\begin{figure}[htbp]
    \centering
        \includegraphics[scale=.35]{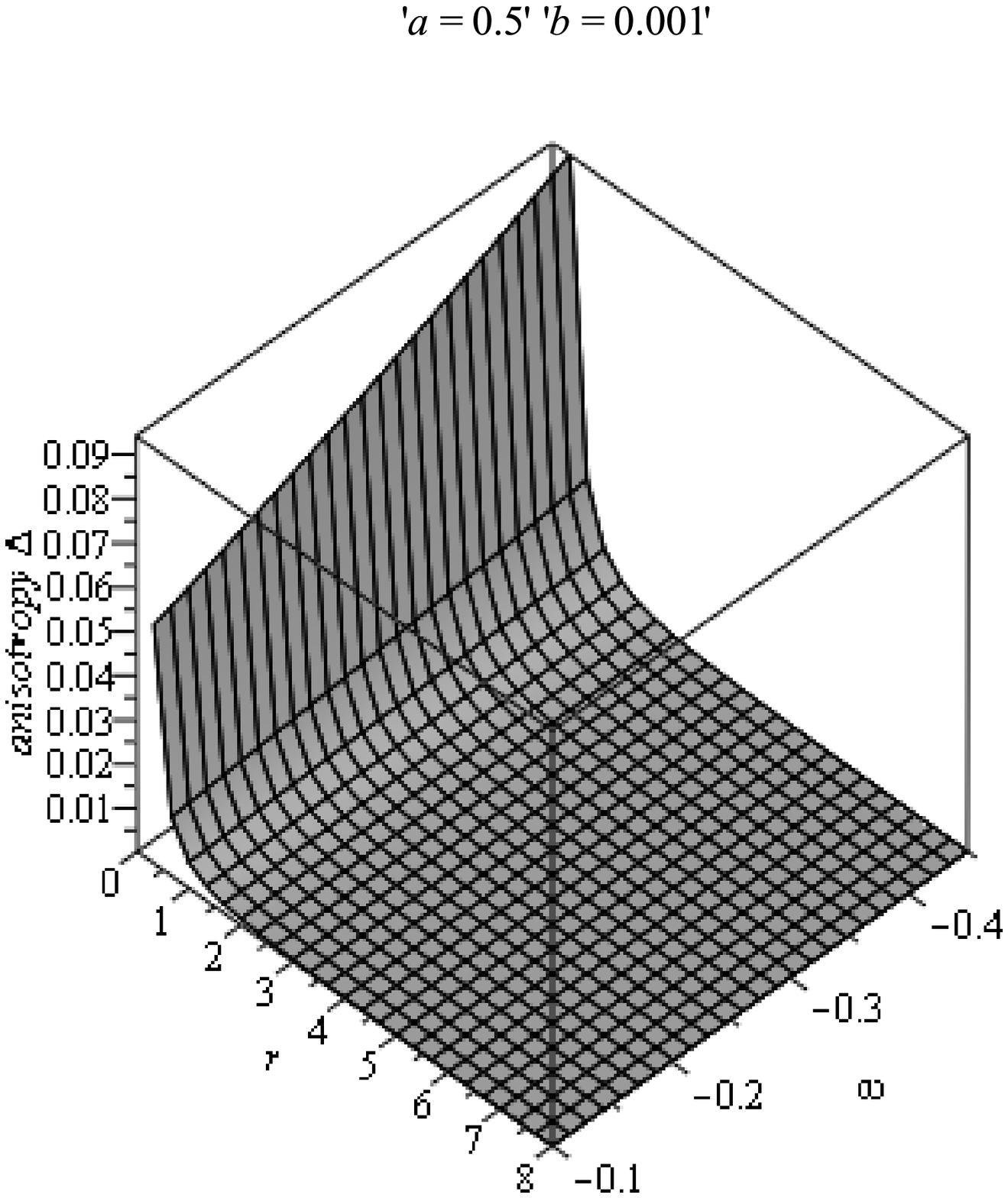}
       \caption{The anisotropy parameter $\Delta=p_t-p_r$ has been shown against
       $r$ for $-0.45\leq \omega\leq -0.1$.}
    \label{fig:3}
\end{figure}

\begin{figure}[htbp]
    \centering
        \includegraphics[scale=.35]{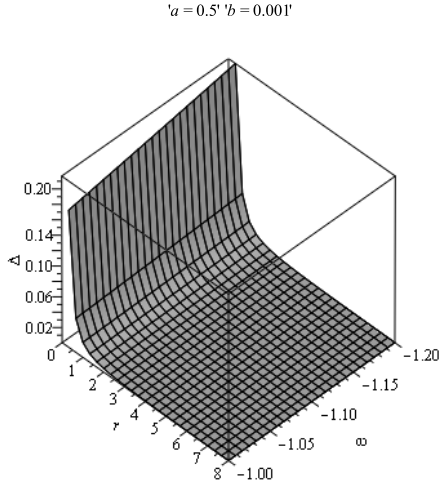}
       \caption{The anisotropy parameter $\Delta=p_t-p_r$ has been
        shown against $r$ for $\omega<-1$}
    \label{fig:3}
\end{figure}

   Now, one can notice that there is a problem in the model such as
    the divergence of the physical
   quantities (energy density and pressures) in the origin. Now to
   overcome this problem in order to model a star, we propose that
   the star contains a core   up to radius $r_1$.

\subsection{Core Solution}

 To avoid this central singularity we cut the spacetime (1) around its origin and placed
 an anisotropy fluid of constant density $\rho_0$ (say).\\
Now, to find the core solution, we assume the radial equation of
state as,
\begin{equation}
p_r=k \rho_0,~~~~k<0
\end{equation}
Here,  the mass function becomes,
\begin{equation}
m(r)=Br^{3}
\end{equation}
where $B=\frac{4\pi \rho_0}{3}$. \\

 Using this expression of
$m(r)$ from equation (4) and  using equation (14)   we get,
\begin{equation}
g(r)=\frac{Br(1+3k)}{1-2Br^{2}}
\end{equation}
Therefore the spacetime metric of the core is given by
\begin{equation}
ds^{2}=(1-2Br^{2})^{-\frac{1+3k}{2}}dt^{2}+\frac{dr^{2}}{1-2Br^{2}}+r^{2}(d\theta^{2}
+\sin^{2}\theta
d\phi^{2})
\end{equation}
The transverse pressure can be obtained as
\begin{equation}
p_t=k\rho_0\left[1+\frac{(1+k)(1+3k)Br^{2}}{2k(1-2Br^{2})}\right]
\end{equation}
and the anisotropic factor $\Delta$ can be obtained as,
\begin{equation}
\Delta=\frac{3}{8\pi}(1+k)(1+3k)\frac{B^{2}r^{2}}{1-2Br^{2}}
\end{equation}
From the expression of $\Delta$ it is clear that $\Delta>0$ if
$k<-1$ and $\Delta<0$ if $-1<k<-\frac{1}{3}$. At the center of
the star $\Delta=0$
 which is expected for a physically reasonable solution. It can also be
 noted that for $k=-1$ and $k=-\frac{1}{3}$ the anisotropic pressure of the
 core reduces to isotropic pressure.  \\

\section{Energy Conditions}

Our particular model of dark energy star consists of five regions:
\begin{enumerate}
                \item Solid core of constant matter density
                \item Thin shell between core and Interior
                \item Interior
                \item Thin shell between interior and exterior spacetime
                \item Exterior Schwarzschild spacetime
              \end{enumerate}
To check whether  our model satisfies all the energy conditions, we have to verify the energy conditions in first four regions described earlier.\\

At first check all  the energy conditions,
    for regions 1 and 3 and later we will discuss the region 2.  We will consider the following inequalities:
\begin{equation}
(i)~NEC:\rho+p_r\geq0
\end{equation}
\begin{equation}
(ii)~WEC:\rho+p_r\geq0,~~\rho\geq0
\end{equation}
\begin{equation}
(iii)~SEC: \rho+p_r\geq0,~~~ \rho+p_r+2p_t\geq0
\end{equation}
\begin{equation}
(iv)~DEC:\rho>\left|p_r\right|,~~~\rho>\left|p_t\right|
\end{equation}
The  fig:6 indicates that for our model,
  all the energy conditions are satisfied in the interior region. The fig:7, however, shows
the SEC is violated within the core.

\begin{figure}[htbp]
    \centering
        \includegraphics[scale=.3]{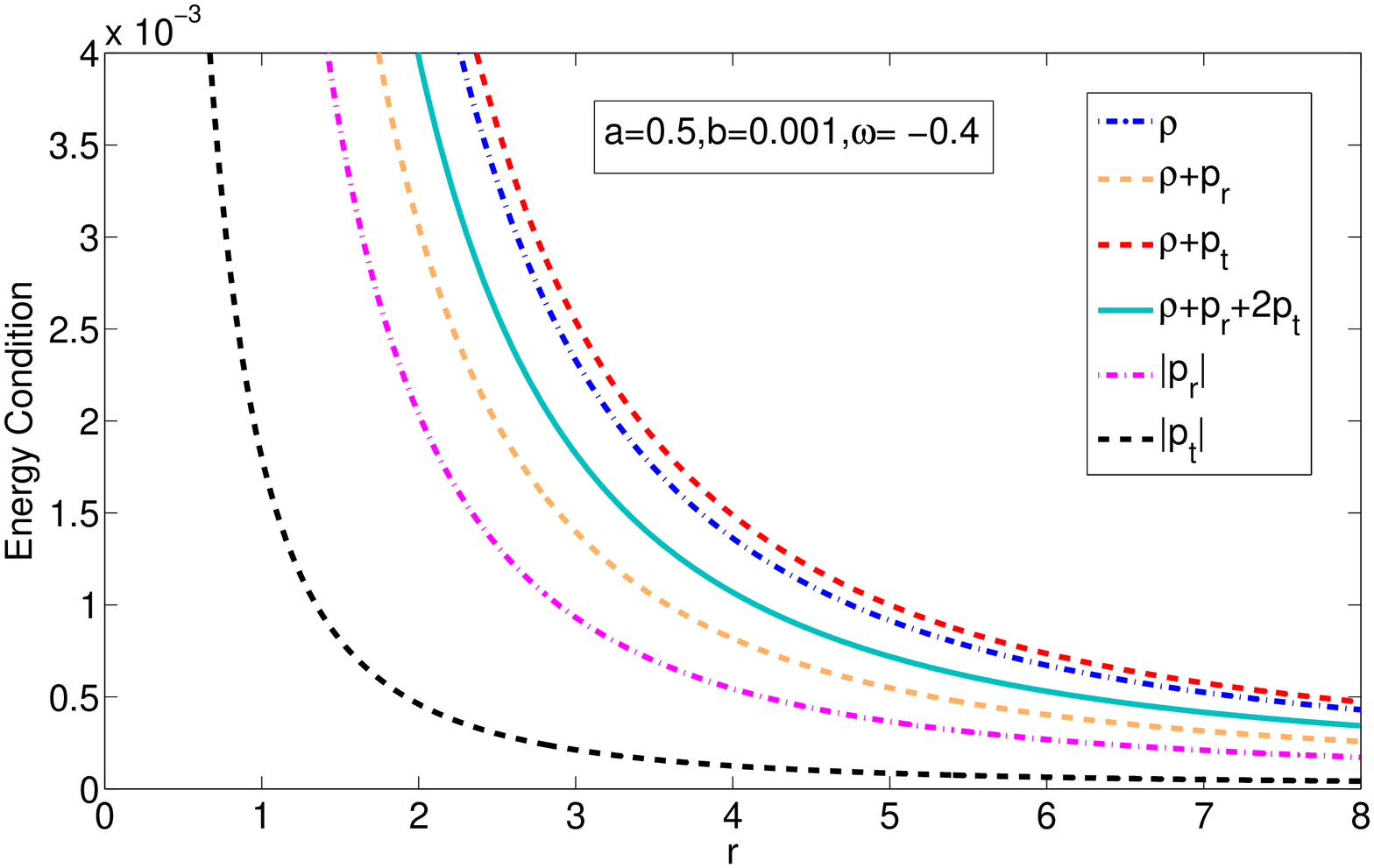}
       \caption{The energy conditions in the interior has been plotted against $r$}
    \label{fig:3}
\end{figure}

\begin{figure}[htbp]
    \centering
        \includegraphics[scale=.35  ]{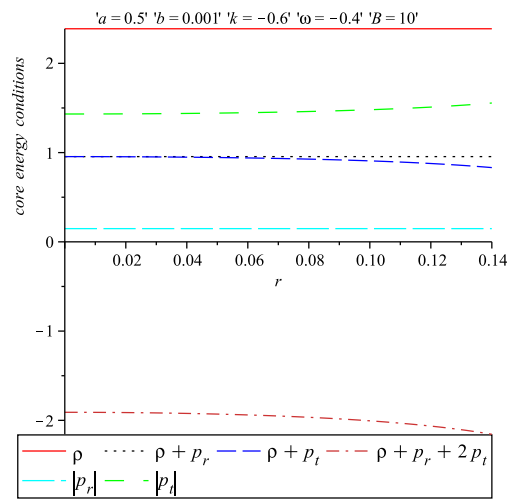}
       \caption{The energy condition of within the core has been plotted against $r$.}
    \label{fig:3}
\end{figure}

\pagebreak

\section{Exterior Spacetime and Junction Condition}

In this section we match our interior spacetime to the exterior schwarzschild vacuum solution along the junction surface with the junction radius $'R'$.The exterior spacetime is given by the metric
\begin{equation}
ds^{2}=-\left(1-\frac{2M}{r}\right)dt^{2}+\frac{dr^{2}}{1-\frac{2M}{r}}+r^{2}(d\theta^{2}+\sin^{2}\theta d\phi^{2})
\end{equation}
Here,event horizon lies at $r=2M$.So obviously $R>2M$. \\

Previously, we have matched our interior spacetime to the exterior Schwarzschild at the boundary $r=R$. Obviously the metric coefficients are continuous at $r=R$, but it does not ensure that their derivatives are also continuous at the junction surface. In other words the affine connections may be discontinuous there. To take care of this let us use the Darmois-Israel\cite{isreal1,isreal2} formation to determine the surface stresses at the junction boundary. The intrinsic surface stress energy tensor $S_{ij}$ is given by Lancozs equations in the following form
\begin{equation}
S^{i}_{j}=-\frac{1}{8\pi}(\kappa^{i}_j-\delta^{i}_j\kappa^{k}_k)
\end{equation}
The discontinuity in the second fundamental form is given by,
\begin{equation}
K_{ij}=K_{ij}^{+}-K_{ij}^{-}
\end{equation}
where the second fundamental form is given by,
\begin{equation}
K_{ij}^{\pm}=-n_{\nu}^{\pm}\left[\frac{\partial^{2}X_{\nu}}{\partial \xi^{1}\partial\xi^{j}}+\Gamma_{\alpha\beta}^{\nu}\frac{\partial X^{\alpha}}{\partial \xi^{i}}\frac{\partial X^{\beta}}{\partial \xi^{j}} \right]|_S,
\end{equation}
where $n_{\nu}^{\pm}$ are the unit normal vector defined by,
\begin{equation}
n_{\nu}^{\pm}=\pm\left|g^{\alpha\beta}\frac{\partial f}{\partial
X^{\alpha}} \frac{\partial f}{\partial X^{\beta}}
\right|^{-\frac{1}{2}}\frac{\partial f}{\partial X^{\nu}},
\end{equation}
with $n^{\nu}n_{\nu}=1$. Here $\xi^{i}$ is the intrinsic coordinate
 on the shell.$+$and$-$ corresponds
 to exterior i.e., Schwarzschild spacetime and interior(our) spacetime
  respectively. The model of our dark energy star is consisting with
   five  zones namely, core, the thin shell between core and interior,
    an inhomogeneous interior region with
anisotropic pressures, thin shell and the exterior vacuum region. \\

Considering the spherical symmetry of the spacetime surface stress energy tensor can
 be written as $S^{i}_j=diag(-\sigma,\mathcal{P},\mathcal{P})$, where $\sigma$ and
 $\mathcal{P}$ are the surface energy density and surface pressure respectively.
\begin{equation}
K_{\tau}^{\tau+}=\frac{\frac{M}{R^{2}}+\ddot{R}}{\sqrt{1-\frac{2M}{R}+\dot{R}^{2}}}
\end{equation}
\begin{equation}
K_{\tau}^{\tau-}=\frac{\frac{(1+\omega)a+bR^{2}(1+3\omega)}{2R}+\ddot{R}-
\frac{(1+\omega)\dot{R}^{2}(a+3bR^{2})}{2(R-Ra-bR^{3})}}{\sqrt{1-a-bR^{2}+\dot{R}^{2}}}
\end{equation}

\begin{equation}
K_{\theta}^{\theta+}=\frac{1}{R}\sqrt{1-\frac{2M}{R}+\dot{R}^{2}}
\end{equation}
\begin{equation}
K_{\theta}^{\theta-}=\frac{1}{R}\sqrt{1-a-bR^{2}+\dot{R}^{2}}
\end{equation}

 The expressions of $\sigma$ and $\mathcal{P}$ are given by,
\begin{equation}
\sigma =-\frac{1}{4\pi R}\left[\sqrt{1-\frac{2M}{R}+\dot{R}^{2}}-\sqrt{1-(a+bR^{2})+\dot{R}^{2}}\right]
\end{equation}
\begin{widetext}
\begin{equation}
\mathcal{P}=\frac{1}{8\pi R}\left[\frac{1-\frac{M}{R}+\dot{R}^{2}+R\ddot{R}}
{\sqrt{1-\frac{2M}{R}+\dot{R}^{2}}}+\frac{1+Rg(R)(1-a-bR^{2}+\dot{R}^{2})
+R\ddot{R}+\frac{bR^{2}}{1-a-bR^{2}}\dot{R}^{2}}{\sqrt{1-a-bR^{2}+\dot{R}^{2}}}\right]
\end{equation}
\end{widetext}

Using conservation identity $S_{j,i}^{i}=-[\dot{\sigma}+2\frac{\dot{R}}{R}(\mathcal{P}+\sigma)]$, one can obtain
\begin{equation}
\sigma'=-\frac{2}{R}(\mathcal{P}+\sigma)+\Xi
\end{equation}
where $\Xi$ is given by,
\begin{equation}
\Xi=-\frac{1}{4\pi R}\frac{m-m'R}{R-2m}\sqrt{1-a-bR^{2}+\dot{R}^{2}}
\end{equation}
The surface mass of the thin shell is given by
\begin{equation}
m_s=4\pi R^{2}\sigma
\end{equation}
Using the expression of $\sigma$ given in equation $(22)$ (considering the static case) we get,
\begin{equation}
m_s=R\left[\sqrt{1-(a+bR^{2})}-\sqrt{1-\frac{2M}{R}}\right]
\end{equation}
After some little manipulation of equation $(27)$ the total mass of the dark energy
star can be obtained as,
\begin{equation}
M=\frac{1}{2}R(a+bR^{2})-\frac{m_s^{2}}{2R}+m_s\sqrt{1-(a+bR^{2})}
\end{equation}
From equation $(27)$ one can obtain
\begin{equation}
\left(\frac{m_s}{2R}\right)''=\Upsilon-4\pi \sigma'\eta
\end{equation}
(for details calculation see the {\bf appendix})
where,
\begin{equation}
\eta=\frac{\mathcal{P'}}{\sigma'},~~~~~~~~\Upsilon=\frac{4\pi}{R}(\sigma+\mathcal{P})+2\pi R\Xi'
\end{equation}
where the 'prime' denotes derivative with respect to 'R'.\\

 ~~We will use the parameter $\eta$ to discuss the stability analysis of the system.
 This $\sqrt{\eta}$ is generally interpreted as the velocity of the sound. So,
 for the physical acceptability one must have $0<\eta \leq 1$. The profile of $\eta$
 has been shown in   $fig.8$ and $fig.9$. \\

%%%%%%%%%%%%%%%%%%%%%%%%%%%%%%%%%%%%%%%%%%%%%%%%%%%%%%%%%%%%%%%%%%%%%%%%%%%%%%%%%%%%%%%%%%%%%
Next we will discuss about the evolution identity given by,
\begin{equation}
\left[T_{\mu\nu}n^{\mu}n^{\nu}\right]_{-}^{+}=\bar{K}_j^{i}S_i^{j}
\end{equation}
where $\bar{K}_j^{i}=\frac{1}{2}\left(K_j^{i+}+K_j^{i-}\right)$
From equation $(31)$ using the equation $(18)-(21)$ one can obtain
\begin{widetext}
\[p_r+\frac{(\rho+p_r)\dot{R}^{2}}{1-a-bR^{2}}=-\frac{1}{2R}\left(\sqrt{1-\frac{2M}{R}+\dot{R}^{2}}+\sqrt{1-a-bR^{2}
+\dot{R}^{2}}\right)\mathcal{P}\]
\begin{equation}
+\frac{1}{2}\left(\frac{\frac{M}{R^{2}}+\ddot{R}}{\sqrt{1-\frac{2M}{R}+\dot{R}^{2}}}+\frac{\frac{(1+\omega)a+bR^{2}(1+3\omega)}{2R}+\ddot{R}
-\frac{(1+\omega)\dot{R}^{2}(a+3bR^{2})}{2(R-Ra-bR^{3})}}{\sqrt{1-a-bR^{2}+\dot{R}^{2}}}\right)\sigma
\end{equation}
\end{widetext}

 Considering static solution at $R=R_0$ with $\dot{R} = \ddot{R}=0$,  we get,
 \begin{widetext}
\begin{equation} p_r =-\frac{1}{2R_0}\left(\sqrt{1-\frac{2M}{R_0} }+\sqrt{1-a-bR_0^{2}
 }\right)\mathcal{P}
+\frac{1}{2}\left(\frac{\frac{M}{R_0^{2}} }{\sqrt{1-\frac{2M}{R_0} }}+\frac{\frac{(1+\omega)a+bR_0^{2}(1+3\omega)}{2R_0}
 }{\sqrt{1-a-bR_0^{2} }}\right)\sigma
\end{equation}
\end{widetext}
Here, $\sigma <0$, therefore,  $p_r<0,$ i.e. tension is in radial  direction.
 Hence a positive tangential surface pressure $P>0$ is required to keep the shell
  stable i.e.  to hold the shell against collapsing.
%%%%%%%%%%%%%%%%%%%%%%%%%%%%%%%%%%%%%%%%%%%%%%%%%%%%%%%%%%%%%%%%%%%%%%%%%%%%%%%%%%%%%%%%%%%%%%%%%%%%%%%%%%%%%%%
\begin{figure}[htbp]
    \centering
        \includegraphics[scale=.25]{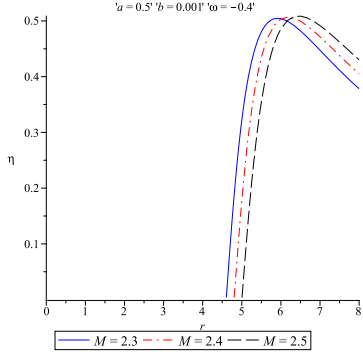}
       \caption{The velocity of the sound  for different values of mass
        $M$ has been plotted against $r$ when the dark energy parameter $\omega$ is fixed }
    \label{fig:3}
\end{figure}
\begin{figure}[htbp]
    \centering
        \includegraphics[scale=.35]{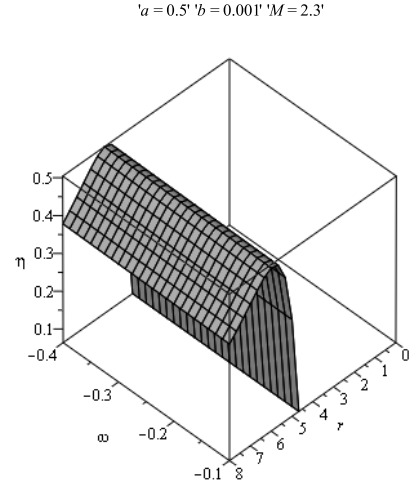}
       \caption{The velocity of the sound  for different values of
         energy parameter $\omega$  has been plotted against $r$ when
          the mass of the dark energy star $M$ is fixed}
    \label{fig:3}
\end{figure}

\subsection{Junction conditions between core  and interior solution }

Since  core  radius  is $r_1$, therefore, we match the core
solution with  interior solution at the junction surface with
junction radius $r_1$.

 The core extrinsic curvature is given by,
\[K_{\tau\tau}^{-}=-Br_1(1+3k)(1-2Br_1^{2})^{-\frac{1}{2}}\]
\[K_{\theta\theta}^{-}=r_1(1-2Br_1^{2})^{\frac{1}{2}}\]
Therefore the surface energy density and surface pressure can be
obtained as,
\begin{equation}
\sigma=-\frac{1}{4\pi
r_1}\left[\sqrt{1-a-br_1^{2}}-\sqrt{1-2Br_1^{2}}\right]
\end{equation}
\begin{equation}
\mathcal{P}=\frac{1}{8\pi r_1}\left[\frac{1+r_1
g(r_1)(1-a-br_1^{2})}{\sqrt{1-a-br_1^{2}}}-\frac{1+(1+3k)Br_1^{2}}{\sqrt{1-2Br_1^{2}}}\right]
\end{equation}

\subsection{ Energy conditions for both the thin shells i.e.
 Interior thin shell between core and interior region and Outer thin shell
  between the interior region and the Schwarzschild spacetime}

According to \cite{hawking} for a thin shell all the energy conditions namely
 Null Energy conditions (NEC), Weak Energy Conditions (WEC), Dominant Energy
 conditions (DEC), Strong Energy Conditions (SEC) will be satisfied if the following
  inequalities hold.\\

\begin{enumerate}
  \item  NEC~:~$\sigma+\mathcal{P} \geq 0$
  \item  WEC~:~$\sigma \geq 0$ and $\sigma+\mathcal{P} \geq 0$
  \item  SEC~:~$\sigma +\mathcal{P}\geq 0$ and $\sigma+2\mathcal{P} \geq 0$
  \item  DEC~:~$\sigma+\mathcal{P} \geq 0$ and $\sigma-\mathcal{P} \geq 0$
\end{enumerate}

The $fig.10$ and $fig.11$ show all energy conditions except DEC,
is violated within both interior and exterior thin shells.

 \begin{figure}[htbp]
     \centering
      \includegraphics[scale=.35]{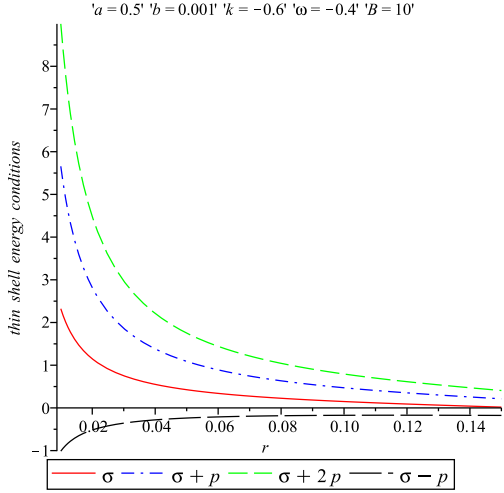}
       \caption{ The energy condition of the thin shell between core and interior region has been plotted against $r$.}
    \label{fig:3}
 \end{figure}

\begin{figure}[htbp]
     \centering
      \includegraphics[scale=.35]{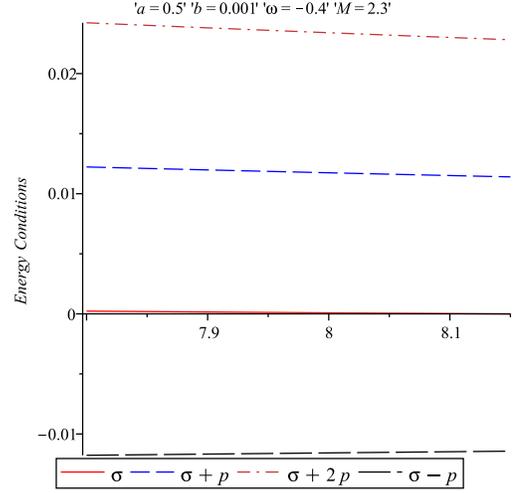}
       \caption{ The energy condition of the thin shell between interior region and the exterior Schwarzschild spacetime has been plotted against $r$.}
    \label{fig:3}
 \end{figure}

\section{TOV Equation}
The generalized Tolman-Oppenheimer-Volkov (TOV) equation is given by the equation \cite{leon}
\begin{equation}
-\frac{M_G(\rho+p_r)}{r^{2}}e^{\frac{\lambda-\nu}{2}}-\frac{dp_r}{dr}+\frac{2}{r}(p_t-p_r)=0
\end{equation}
Where $M_G=M_G(r)$ is the effective gravitational mass inside a sphere of radius $r$
 given by the Tolmam-Whittaker formula which can be derived from the equation
\begin{equation}
M_G(r)=\frac{1}{2}r^{2}e^{\frac{\nu-\lambda}{2}}\nu'
\end{equation}
The above equations describes the equilibrium conditions of the fluid sphere
subject to gravitational,hydrostatics and anisotropy forces.
\begin{figure}[htbp]
    \centering
        \includegraphics[scale=.27]{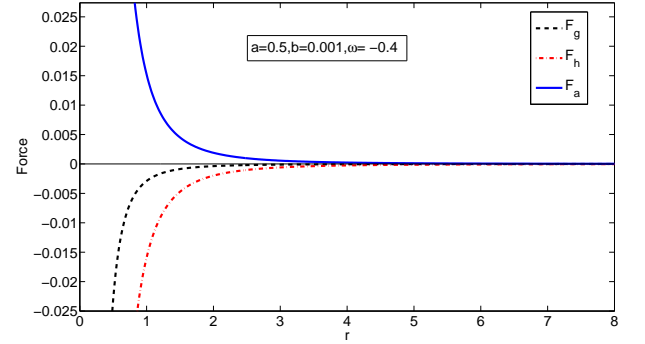}
       \caption{The dark energy star is in static equilibrium under gravitational
       $(F_g)$,hydrostatics $(F_h)$ and anisotropy $(F_a)$ forces.}
    \label{fig:3}
\end{figure}

The equation $(34)$ can be modified in the form
\begin{equation}
F_g+F_h+F_a=0
\end{equation}
where
\begin{equation}
F_g=-\frac{\nu'}{2}(\rho+p_r)
\end{equation}
\begin{equation}
F_h=-\frac{dp_r}{dr}
\end{equation}
\begin{equation}
F_a=\frac{2}{r}(p_t-p_r)
\end{equation}
The profiles of $F_g,F_h,F_a$ has shown in {  $fig.12$}.
 The figure shows that our dark energy model is in static equilibrium under
  gravitational $(F_g)$ , hydrostatics $(F_h)$ and anisotropic $(F_a)$ forces.

\section{Mass radius relation}

The mass of the dark energy star has been given in equation $(8)$.\\
The compactness of the star is defined as
\begin{equation}
u=\frac{m(r)}{r}=\frac{1}{2}(a+br^{2})
\end{equation}
and the surface redshift is defined by
\begin{equation}
Z_s=(1-2u)^{-\frac{1}{2}}-1=(1-a-br^{2})^{-\frac{1}{2}}-1
\end{equation}
The profile of mass function,compactness and surface redshift of the
 dark energy star have been given in  $ figs.13,~ 14 ~and~ 15$  respectively.\\
\begin{figure}[htbp]
    \centering
        \includegraphics[scale=.3]{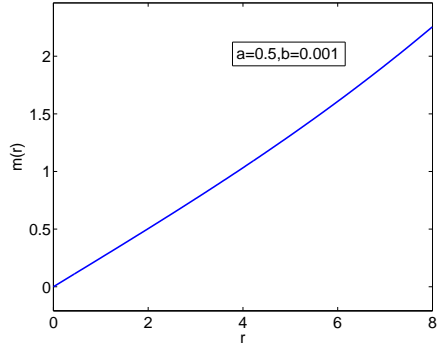}
       \caption{mass function $m(r)$ has been shown against $r$.}
    \label{fig:3}
    \end{figure}

    \begin{figure}[htbp]
    \centering
        \includegraphics[scale=.3]{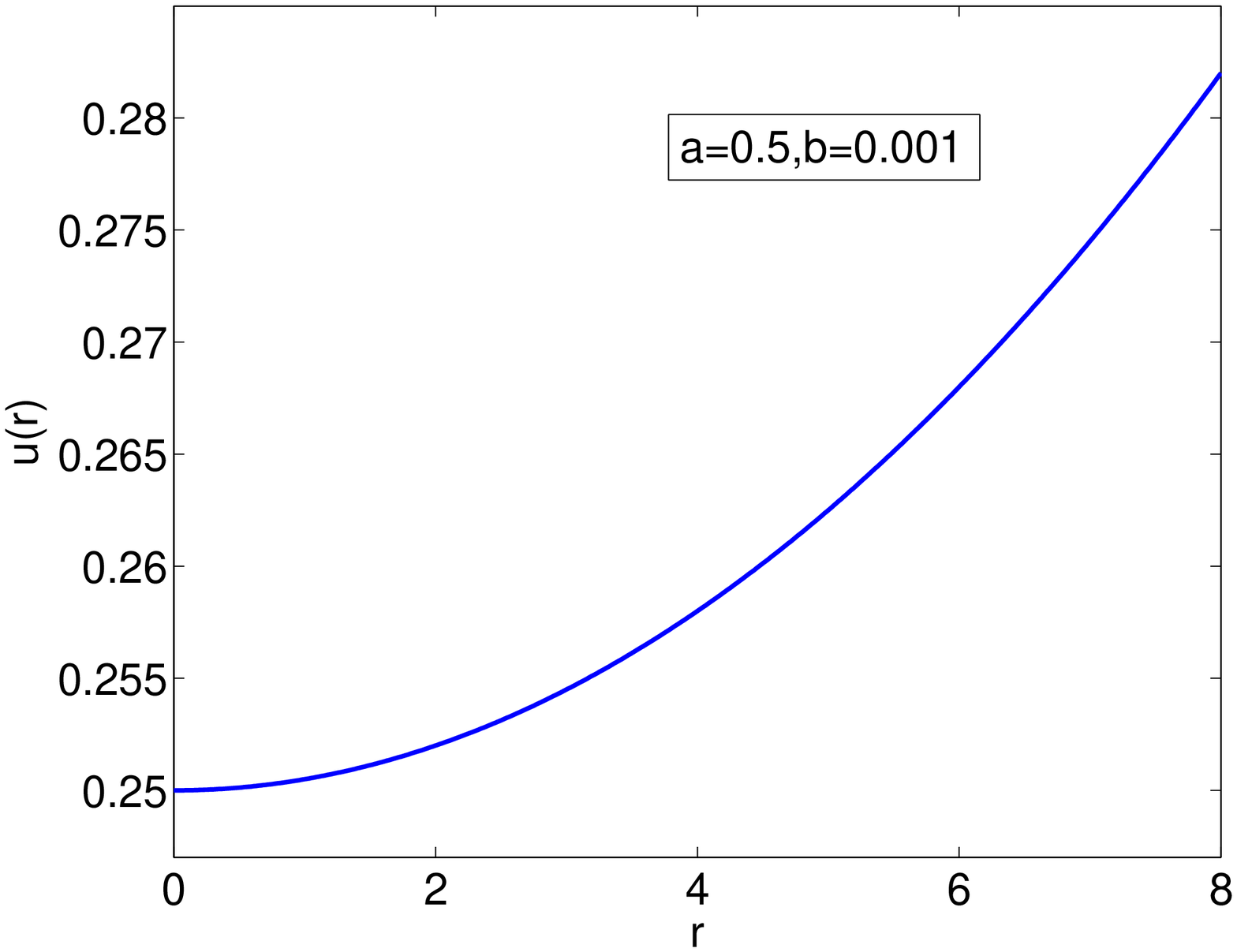}
       \caption{The compactness of the dark energy star has been shown against $r$.}
    \label{fig:3}
\end{figure}

\begin{figure}[htbp]
    \centering
        \includegraphics[scale=.3]{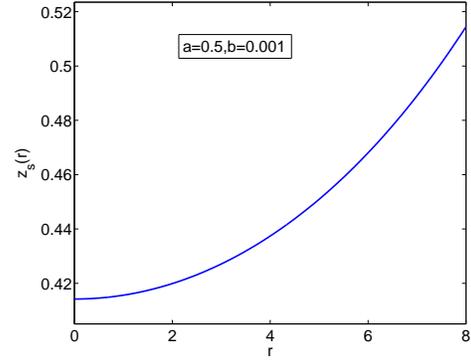}
       \caption{Surface redshift $ Z_s$ has been shown against $r$}
    \label{fig:3}
\end{figure}

\section{Stability Analysis}
In this section we are going to analyzed the stability of our model.\\
Rearranging the equation $(22)$ we get,
\begin{equation}
\dot{R}^{2}+V(R)=0
\end{equation}
Where $V(R)$ is given by,
\begin{equation}
V(R)=1-\frac{M-m}{R}-\left(\frac{m_s}{2R}\right)^{2}-\left(\frac{M-m}{m_s}\right)^{2}
\end{equation}

(For details derivation see  {Appendix:1})\\

To discuss the linearized stability analysis let us take a linear perturbation around a static radius $R_0$. Expanding $V(R)$ by Taylor series around the radius of the static solution $ R= R_0$ one can obtain

\[V(R)=V(R_0)+(R-R_0)V'(R_0)+\frac{(R-R_0)^{2}}{2}V''(R_0)\]
\begin{equation}~~~~~~~~~~~~~~~~~~~~~~~~~~~+O[(R-R_0)^{3}]
\end{equation}

where 'prime' denotes derivative with respect to $'R'$ \\
Since we are linearizing around static radius $R=R_0$ we must have $V(R_0)=0,V'(R_0)=0$.The configuration will be stable if $V(R)$ has a local minimum at $R_0$ i.e,if  $V''(R_0)>0$ \\
Now from the relation $ V'(R_0)=0$ we get,
\begin{equation}
\left(\frac{m_s(R_0)}{2R_0}\right)'
=A\left[F'(R_0)-2\left(\frac{M-m(R_0)}{m_s}\right)\left(\frac{M-m(R_0)}{m_s}\right)'\right]
\end{equation}
where A is given in (*)\\
Now the configuration will be stable if $V''(R_0)>0$.i.e if
\begin{equation}
\eta \frac{d}{dR}(\sigma^{2})>\frac{1}{2\pi}\left[\sigma\Upsilon-\frac{1}{2\pi R_0}(H^{2}-G^{2})\right]
\end{equation}

For details derivation see  {appendix:3}

\begin{figure}[htbp]
    \centering
        \includegraphics[scale=.3]{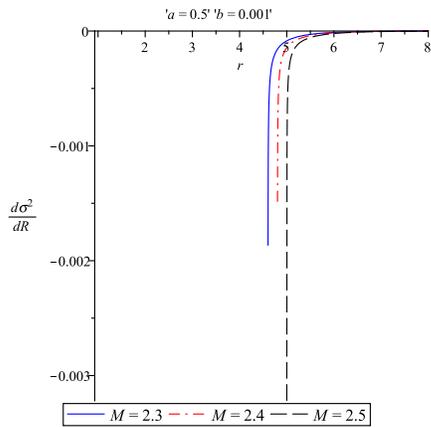}
       \caption{$\frac{d\sigma^{2}}{dR}$ has been shown against $R$}
    \label{fig:3}
\end{figure}

\begin{equation}
G(R_0)=A\left[F'(R_0)-2\left(\frac{M-m(R_0)}{m_s(R_0)}\right)
\left(\frac{M-m(R_0)}{m_s(R_0)}\right)'\right]
\end{equation}
$$A=\left(\frac{R_0}{m_s(R_0)}\right) $$

\[\left\{H(R_0)\right\}^{2}=\frac{1}{2}F''(R_0)-\left[\frac{M-m(R_0)}{m_s(R_0)}\right]^{2}\]
\begin{equation}
-\left(\frac{M-m(R_0)}{m_s(R_0)}\right)\left(\frac{M-m(R_0)}{m_s(R_0)}\right)''
\end{equation}

Now from equation $(50)$ we get
\begin{equation}
\eta_0\frac{d\sigma^{2}}{dR}|_{R_0}>\Omega
\end{equation}
where $\Omega=\frac{1}{2\pi}\left[\sigma\Upsilon-\frac{1}{2\pi R_0}(H^{2}-G^{2})\right]$.From equation $(53)$ the stability regions are dictated by the following inequalities
\begin{equation}
\eta_0>\Omega \left(\frac{d\sigma^{2}}{dR}|_{R_0}\right)^{-1} ~~~if \frac{d\sigma^{2}}{dR}|_{R_0} >0
\end{equation}

\begin{equation}
\eta_0<\Omega\left(\frac{d\sigma^{2}}{dR}|_{R_0}\right)^{-1} ~~~if \frac{d\sigma^{2}}{dR}|_{R_0} < 0
\end{equation}
From the plot of $\frac{d\sigma^{2}}{dR}$ (see   $fig. 16 $ ) we
see that $\frac{d\sigma^{2}}{dR}<0$. So the stability region for
our model is given by equation $(55)$.

\section{Discussions and concluding remarks }

In this work we have obtained a new class of exact interior solution by choosing
 a special form of energy density which describes a model of dark energy star
 parameterized by $\omega=\frac{p_r}{\rho}<0$.   For our choice of $a=0.5$ and
$b=0.001$, we have shown that dark energy parameter $\omega$ lies
in either
  $-0.5<\omega<-0.1$ or $\omega<-1.$ The obtained solutions are well behaved for
  $r>0$. From the figures 1 and 2,  we see that gravity profile $ g(r)>0 $ when
  $-0.5\leq \omega\leq -0.1$   and $g(r)<0$ when $\omega$ lies in the phantom regime.
  The energy density $\rho$, radial pressure $(p_r)$, transverse pressure $(p_t)$ all
  are monotonic decreasing function of $r$. The anisotropy factor $\Delta>0$ for
   $-0.5<\omega<-0.1$ as well as for $\omega<-1$ which implies $p_t>p_r$ i.e.
    the anisotropic force is attractive in nature.
      We have matched our interior spacetime to the exterior
Schwarzschild spacetime in presence of thin shell where we have assumed positive
surface pressure to hold the thin shell against collapse. The mass of the dark energy
 star in terms of the thin shell mass has been proposed as well as the relationship
  among $p_r, \sigma,\mathcal{P}$ has been given. By keeping $\omega$ fixed and choosing
   different values of of $M$, we have shown that $0<\eta<1$. Similarly by keeping
    the mass $M$ fixed and for $-0.45\leq \omega<-0.1 $, we have shown that $0<\eta<1$.
     All the energy conditions in the interior region are   satisfied.
     However,
   in the core SEC is violated  and both the thin shells i.e.
 interior thin shell between core and interior region and Outer thin shell
  between the interior region and the Schwarzschild spacetime, DEC is violated.

     The mass function
     is monotonic increasing and regular at the center. For $(3+1) $ dimensional astrophysical
      object, Buchdahl\cite{buch} has shown that $\frac{2M}{R}<\frac{8}{9}$.
       For our model $\frac{2M}{R}=0.564<\frac{8}{9}$. The stability analysis
       under small radial perturbation has also been discussed. \\

\section{Acknowledgements}

  FR gratefully acknowledge support from the Inter-University Centre for Astronomy and
  Astrophysics (IUCAA),
Pune, India.  We are very grateful   to an anonymous referee for
his/her insightful comments that have led to significant
improvements, particularly on the interpretational
aspects.\\

\pagebreak

\pagebreak

 {\Large{\textbf{Appendix.1}}}

\[m_s=4\pi R^{2}\sigma\]
using the expression of $\sigma$ we get,
\[or~~\frac{m_s}{4\pi R^{2}}=\frac{1}{4\pi R}\left[\sqrt{1-\frac{2m}{R}+\dot{R}^{2}}-\sqrt{1-\frac{2M}{R}+\dot{R}^{2}}\right]\]
\[or,~~\frac{m_s}{a}=\sqrt{1-\frac{2m}{R}+\dot{R}^{2}}-\sqrt{1-\frac{2M}{R}+\dot{R}^{2}}\]
\[or~~~\frac{m_s}{a}-\sqrt{1-\frac{2m}{R}}=-\sqrt{1-\frac{2M}{R}+\dot{R}^{2}}\]
Squaring bothside we get,
\[\left(\frac{m_s}{R}\right)^{2}-2\frac{m_s}{R}\sqrt{1-\frac{2m}{R}+\dot{R}^{2}}=\frac{2}{R}(m-M)\]
\[or,~~~~~~~\frac{m_s}{R}\left[\frac{m_s}{R}-2\sqrt{1-\frac{2m}{R}+\dot{R}^{2}}\right]=\frac{2}{R}(m-M)\]
\[or,~~~\frac{m_s}{R}-2\sqrt{1-\frac{2m}{R}+\dot{R}^{2}}=\frac{2}{m_s}(m-M)\]
\[or,~~~\frac{m_s}{R}-\frac{2}{m_s}(m-M)=2\sqrt{1-\frac{2m}{R}+\dot{R}^{2}}\]
\[or,~~~~\frac{m_s}{2R}+\frac{M-m}{m_s}=\sqrt{1-\frac{2m}{R}+\dot{R}^{2}}\]
again squaring bothside we get,
\[\left(\frac{m_s}{2R}\right)^{2}+\left(\frac{M-m}{m_s}\right)^{2}+2\frac{M-m}{2R}=1-\frac{2m}{R}+\dot{R}^{2}\]
which gives,
\[\dot{R}^{2}=\left(\frac{m_s}{2R}\right)^{2}+\left(\frac{M-m}{m_s}\right)^{2}+\frac{M-m}{R}-1\]
Now,\[\dot{R}^{2}=-V(R)\]
which gives,
\[V(R)=1-\frac{M-m}{R}-\left(\frac{m_s}{2R}\right)^{2}-\left(\frac{M-m}{m_s}\right)^{2}\]
\\

{\Large{\textbf{Appendix.2}}}
\[m_s=4\pi R^{2}\sigma \]
\[or,~~\frac{m_s}{2R}=2\pi R \sigma\]
Differentiating bothside with respect to R we get,
\[or,~~\left(\frac{m_s}{2R}\right)'=2\pi (R \sigma'+\sigma)\]
\[~~~~~~~~~~~~~~~~~~~~~~~~~~~~~~~~~~~~~~=2\pi R\left\{-\frac{2}{R}(\sigma+\mathcal{P})+\Xi \right\}+2\pi \sigma\]
\[~~~~~~~~~~~~~~~~~~~=-4\pi\mathcal{P}+2\pi R \Xi-2\pi\sigma\]
Differentiating bothside with respect to R we get,
\[\left(\frac{m_s}{2R}\right)''=-4\pi \mathcal{P}'+2\pi (R \Xi'+\Xi)-2\pi\sigma'\]
Using the value of $\sigma'$ we get,
\[\left(\frac{m_s}{2R}\right)''=-4\pi \mathcal{P}'+2\pi (R \Xi'+\Xi)-2\pi\left\{-\frac{2}{R}(\sigma+\mathcal{P})+\Xi \right\}\]
\[=\frac{4\pi}{R}(\sigma+\mathcal{P})+2\pi R\Xi'-4\pi\eta\sigma'\]
therefore,
\[\left(\frac{m_s}{2R}\right)''=\Upsilon-4\pi \eta\sigma'\]
where,
\[\Upsilon=\frac{4\pi}{R}(\sigma+\mathcal{P})+2\pi R\Xi'\]
\\

{\Large{\textbf{Appendix.3}}}

\[V(R)=F(R)-\left(\frac{m_s}{2R}\right)^{2}-\left(\frac{M-m}{m_s}\right)^{2}\]
\[V'(R)=F'(R)-2\left(\frac{m_s}{2R}\right)\left(\frac{m_s}{2R}\right)'\]
\[-2\left(\frac{M-m}{m_s}\right)\left(\frac{M-m}{m_s}\right)
'\]

Now,$V'(R_0)=0 $ gives,
\begin{widetext}
\[\left(\frac{m_s(R_0)}{2R_0}\right)'
=\left(\frac{R_0}{m_s(R_0)}\right)\left[F'(R_0)-2\left(\frac{M-m(R_0)}{m_s}\right)\left(\frac{M-m(R_0)}{m_s}\right)'\right]\]
\end{widetext}
\begin{widetext}
\[let,~~~\left(\frac{m_s(R_0)}{2R_0}\right)'=G(R_0)=\left(\frac{R_0}{m_s(R_0)}\right)
\left[F'(R_0)-2\left(\frac{M-m(R_0)}{m_s(R_0)}\right)
\left(\frac{M-m(R_0)}{m_s(R_0)}\right)'\right]\]
\end{widetext}
now
\begin{widetext}
\[V''(R)=F''(R)-2\left[\left(\frac{m_s(R)}{2R}\right)\left(\frac{m_s(R)}{2R}\right)''
+\left\{\left(\frac{m_s(R)}{2R}\right)'\right\}^{2}\right]
-2\left[\left(\frac{M-m(R)}{m_s(R)}\right)
\left(\frac{M-m(R)}{m_s(R)}\right)''+\left\{\left(\frac{M-m(R)}{m_s(R)}\right)'\right\}^{2}\right]\]
\end{widetext}
\begin{widetext}
\[V''(R_0)=F''(R_0)-2\left[\left(\frac{m_s(R_0)}{2R_0}\right)\left(\frac{m_s(R_0)}{2R_0}\right)''
+G(R_0)^{2}\right]-2\left[\left(\frac{M-m(R_0)}{m_s(R_0)}\right)
\left(\frac{M-m(R_0)}{m_s(R_0)}\right)''+\left\{\left(\frac{M-m(R_0)}{m_s(R_0)}\right)'\right\}^{2}\right]\]
\end{widetext}

Now $V''(R_0)>0$ gives,
\[H(R_0)^{2}>
[G(R_0)]^{2}+\left(\frac{m_s(R_0)}{2R_0}\right)\left(\frac{m_s(R_0)}{2R_0}\right)''\]
where
\[H(R_0)^{2}=\frac{1}{2}F''(R_0)-\left[\frac{M-m(R_0)}{m_s(R_0)}\right]^{2}\]
\[~~~~~~~~~~~~-\left(\frac{M-m(R_0)}{m_s(R_0)}\right)\left(\frac{M-m(R_0)}{m_s(R_0)}\right)''\]

\[or,~~~H^{2}-G^{2}>2\pi R_0\sigma\left[\frac{4\pi}{R_0}(\sigma+\mathcal{P})+2\pi R_0\Xi'-4\pi \eta\sigma'\right]\]
\[or,~~~H^{2}-G^{2} >2\pi R_0\sigma (\Upsilon-4\pi \eta \sigma')\]
\[or,~~~\frac{1}{2\pi R_0}(H^{2}-G^{2})>\sigma\Upsilon-2\pi\eta \frac{d}{da}(\sigma^{2})\]
\[or,~~~~\eta \frac{d}{da}(\sigma^{2})>\frac{1}{2\pi}\left[\sigma\Upsilon-\frac{1}{2\pi R_0}(H^{2}-G^{2})\right]\]


\begin{thebibliography}{99}

\bibitem{yanjun} YAN Jun,{\it Commun. Theor. Phys} {\bf 52},1016 (2009).
\bibitem{Chan1}R. Chan, M. F. A. da Silva, Jaime F. Villas da Rocha,    Mod.Phys.Lett.A {\bf24},
1137(2009).
\bibitem{Chan2}R. Chan, M. F. A. da Silva, Jaime F. Villas da Rocha,{\it Gen.Rel.Grav.}
{\bf41}1835(2009).
\bibitem{Ghezzi}Cristian R. Ghezzi,    {\it Astrophys.Space Sci.} {\bf333}437(2011).
\bibitem{lobo1}Francisco S. N. Lobo,{\it Class.Quant.Grav.}{\bf 23} 1525 (2006).
\bibitem{lobo2}Francisco S. N. Lobo,    {\it Phys.Rev.D}{\bf75} 024023(2007).
\bibitem{saibal} Saibal Ray , Farook Rahaman, Utpal Mukhopadhyay, Ruby Sarkar,
  {\it Int. J. Theor. Phys} {\bf 50}, 2687 (2011).
\bibitem{jadav}A K Yadav, F Rahaman, and S Ray,  {\it Int. J. Theor. Phys.}{\bf 50} , 871 (2011).
\bibitem{dev}Dev K and Gleiser M, {\it Gen. Relat.Grav}, {\bf 34},1793,(2002).
\bibitem{fr1} F. Rahaman Mubasher Jamil, Ranjan Sharma, Kausik Chakraborty,  {\it Astrophys.Space Sci.} {\bf 330} 249 (2010).
\bibitem{9}C.Misner and H.Zapolsky, {\it Phys.Rev.Lett}  {\bf 12},635 (1964).
\bibitem{isreal1} W. Israel, {\it Nuovo Cimento B} {\bf44} (1966) 1.
\bibitem{isreal2} W. Israel, {\it Nuovo Cimento B} {\bf48} (1967) 463 (Erratum).
\bibitem{leon}J.Ponce de Le\'{o}n, {\it Gen.Relativ.Gravit} {\bf 25}, 1123 (1993).
\bibitem{neil}Neil J. Cornish,  arxiv:gr-qc/9405065
\bibitem{dubravko}Dubravko Horvat, Anja Marunovi\'{c},{\it Class.Quant.Grav},\\
{\bf 30}, 145006 (2013)
\bibitem{wen}Wen-Jie Su,Jun Yan,{\it Can.J.Phys} {\bf 90} ,1279 (2012)
\bibitem{paul}Paul Halpern,Michael Pecorino {\it ISRN Astron.Astrophys},\\
{\bf 2013} 939876 (2013)
\bibitem{vlad}Vladimir Folomeev,Ascar Aringazin,Vladimir \\
   Dzhunushaliev {\it Phys.Rev.D} {\bf 88} 063005 (2013)
\bibitem{jorge}Jorge Ovalle,L\'{a}szl\'{o}.\'{A}.Gergely,Roberto Casadio\\
        arXiv:1405.0252 [gr-qc].
\bibitem{chris}Chris Kouvaris,M. Angeles Perez-Garcia {\it Phys.Rev.D} {\bf 89} 103539 (2014)
\bibitem{bridget}Bridget Bertoni,Ann E. Nelson,Sanjay Redd {\it Phys.Rev. D} {\bf 88} 123505 (2013)
\bibitem{buch}H.A.Buchdahl,{\it Phys.Rev} {\bf 116} ,1027, (1959)
\bibitem{misner} C. W. Misner and  H.  S. Zapolsky, {\it Phys. Rev.Lett.} {\bf12}635 (1964)\\
\bibitem{hawking} S.W.Hawking and G.F.R.Ellis, The Large Scale structure of Spacetime,   Cambridge University Press, Cambridge,(1973)\\







\end{thebibliography}
\end{document}